\documentclass[twocolumn,superscriptaddress,letter]{revtex4}%
\usepackage{amssymb}
\usepackage{color}
\usepackage{graphicx}
\usepackage{dcolumn}
\usepackage{bm}
\usepackage[header,title,page,titletoc]{appendix}
\usepackage{amsmath}
\usepackage{amsfonts}%
\providecommand{\U}[1]{\protect \rule{.1in}{.1in}}

\begin{document}
\title{Topological Mid-gap States of Topological Insulators with Flux-Superlattice}
\author{Ya-Jie Wu}
\author{Jing He}
\author{Su-Peng Kou}
\thanks{Corresponding author}
\email{spkou@bnu.edu.cn}
\affiliation{Department of Physics, Beijing Normal University, Beijing 100875, China}

\begin{abstract}
In this paper based on the Haldane model, we study the topological insulator
with superlattice of $\pi$-fluxes. We find that there exist the mid-gap states
induced by the flux-superlattice. In particular, the mid-gap states have
nontrivial topological properties, including the nonzero Chern number and the
gapless edge states. We derive an effective tight-binding model to describe
the topological mid-gap states and then study the mid-gap states by the
effective tight-binding model. The results can be straightforwardly
generalized to other two dimensional topological insulators with flux-superlattice.

\end{abstract}
\maketitle

\section{Introduction}

Topological insulator (TI) is a novel class of insulators with topological
protected metallic edge states (surface) states.
\cite{Haldane,kane,kane2,zhang}. The TIs are robust against local
perturbations. From the "holographic feature" of the topological ordered
states, people may use the topological defects such as the quantized vortices
with half of magnetic flux $\Phi_{0}=hc/2e$ to probe the non-trivial bulk
topology of the system\cite{Ran,Qi}. The fermionic zero modes localize around
the quantized vortices in TI. For the multi-topological-defect that forms a
superlattice, there may exist mid-gap bands.\emph{ }In Ref.\cite{Assaad}, a
vortex-line on a TI has been discussed and the low energy physics was
described by an effective spin model of the fluxons. An interesting question
arises\emph{ "what's the mid-gap system of topological insulator with }$\pi
$\emph{-fluxes that form a two dimensional superlattice?"}

In this paper, based on the Haldane model, we explore the properties of
mid-gap states in a topological insulator with a triangular flux-superlattice.
For the Haldane model, there exists a zero energy state (the so-called zero
mode) of fermions around each $\pi$-flux. The overlap of different zero modes
around the well separated $\pi$-fluxes leads to mid-gap states inside the band
gap of the parent TI. We use an effective tight-binding model to characterize
the mid-gap states induced by superlattice of $\pi$-fluxes. In particular, we
find that this effective tight-binding model has nontrivial topological
properties and the triangular superlattice of $\pi$-fluxes in TI can be
regarded as an emergent topological insulator. See the illustration of TI with
triangular flux-superlattice in Fig.\textcolor[rgb]{1.00,0.00,0.50}{1}. Since
the effective hopping parameters (the tunneling splitting) are manipulated by
adjusting the distance between $\pi$-fluxes, we can tune the ratio between the
effective nearest-neighbor(NN) and next-nearest-neighbor(NNN) hopping
parameters to control the properties of the mid-gap states. The situation is
similar to a topological superconductor (TSC) with vortex-lattice, of which
the mid-gap states are described by a Majorana lattice model and can be
regarded as a "topological superconductor" on the parent topological
superconductor\cite{v,kou}.

The reminder of this paper is organized as follows. In Sec.II, we introduce
the Haldane model and give a brief discussion on it. In Sec. IIIA,
\begin{figure}[h]
\scalebox{0.42
}{\includegraphics* [1.2in,0.20in][9.2in,5.2in]{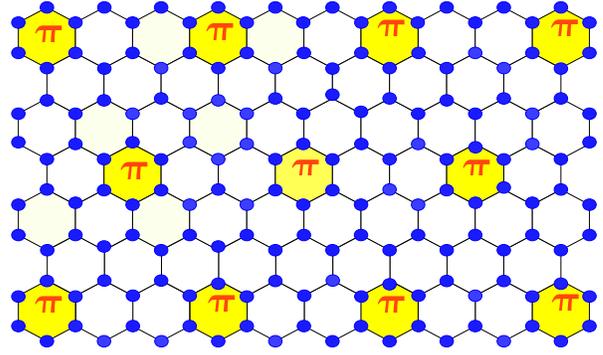}}\caption{The
illustration of the TI (Haldane model in a honeycomb lattice) with triangular
flux-superlattice. Each yellow honeycomb plaquette indicates a $\pi$-flux.}%
\end{figure}we study the properties of TI with two $\pi$-fluxes. In Sec. IIIB,
we investigate the properties of TI with triangular flux-superlattice. In Sec.
IV, we write down an effective tight-binding model to describe the mid-gap
states. Finally, we conclude our discussions in Sec. V.

\section{The Haldane model}

Our starting point is the Haldane model in a honeycomb lattice\cite{Haldane},
of which the Hamiltonian is given by%
\begin{equation}
\hat{H}_{\mathrm{Hal}}=-\mathit{T}\sum_{\left \langle i,j\right \rangle }\hat
{c}_{i}^{\dagger}\hat{c}_{j}-\mathit{T}^{\prime}\sum \limits_{\left \langle
\left \langle {i,j}\right \rangle \right \rangle }e^{i\phi_{ij}}\hat{c}%
_{i}^{\dagger}\hat{c}_{j}-\mu \sum_{i}\hat{c}_{i}^{\dagger}\hat{c}_{i}\text{.}
\label{KM}%
\end{equation}
Here, the fermionic operator $\hat{c}_{i}$ annihilates a fermion on lattice
site $i$. $\mathit{T}$ is the NN hopping amplitude and $\mathit{T}^{\prime}%
$\ is the NNN hopping amplitude. $\left \langle {i,j}\right \rangle $,
$\left \langle \left \langle {i,j}\right \rangle \right \rangle $ denote the NN
and the NNN links, respectively. $e^{i\phi_{ij}}$ is a complex phase along the
NNN link, and we set the direction of the positive phase $\left \vert \phi
_{ij}\right \vert =\frac{\pi}{2}$ clockwise. $\mu$ is the chemical potential
which is set to be zero in this paper. In the following, we take the lattice
constant $a\equiv1$.

The energy spectrums of the free fermions of above Hamiltonian are given by
\begin{equation}
E_{q}=\pm \sum_{q}\sqrt{\xi_{q}^{2}+(\xi_{q}^{\prime})^{2}},
\end{equation}
where
\begin{align}
\xi_{q}  &  =\mathit{T}\left[  3+2\cos{(\sqrt{3}q_{y})}+4\cos{(3q_{x}/2)}%
\cos{(\sqrt{3}q_{y}/2)}\right]  ^{\frac{1}{2}},\nonumber \\
\xi_{q}^{^{\prime}}  &  =2\mathit{T}^{\prime}(\sin(\sqrt{3}q_{y})-2\cos
(3q_{x}/2)\sin{(\sqrt{3}q_{y}/2)}).
\end{align}
From this energy spectrum, there exists the energy gap $\Delta_{f}=6\sqrt
{3}\mathit{T}^{\prime}$ at the Dirac points $\mathbf{q}_{1}=\frac{2\pi}%
{3}(1,1/\sqrt{3})$\ and $\mathbf{q}_{2}=-\frac{2\pi}{3}(1,1/\sqrt{3})$. The
density of states (DOS) of the Haldane model is shown in
Fig.\textcolor[rgb]{1.00,0.00,0.50}{5}(a) for the case of $\mathit{T}^{\prime
}/\mathit{T}=0.1$.

The Haldane model is an integer quantum Hall insulator without Landau levels.
It breaks time reversal symmetry without any net magnetic flux through the
unit cell of a periodic two-dimensional honeycomb lattice.\ There exists
topological invariant for the Haldane model - the TKNN number (or the Chern
number)\cite{dj,Simon}. Thus, the Haldane model in Eq.(\ref{KM}) is a typical
topological band insulator for the case of $\mathit{T}^{\prime}\neq0$ at half-filling.

\section{Mid-gap states of the Haldane model with triangular
flux-superlattice}

In this section, based on the Haldane model described by Eq.(\ref{KM}), we
study the topological insulator with multi-flux, including the two-flux case,
and the triangular flux-superlattice case.

\subsection{Mid-gap states of the Haldane model with two fluxes}

\begin{figure}[ptb]
\centering \scalebox{0.45
}{\includegraphics* [0.2in,0.70in][8.2in,3.2in]{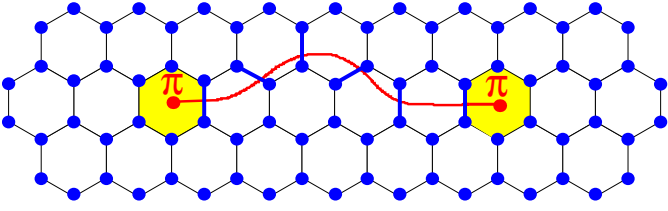}}\caption{The
illustration of the two $\pi$-flux (yellow plaquette) on the honeycomb
lattice. When the fermion hops across the red strings which connect two
fluxes, the wave function changes sign.}%
\end{figure}

Firstly, we consider the TI with two well separated $\pi$-fluxes (for example,
the distance between two fluxes is $5\sqrt{3}a$). See the illustration in
Fig.\textcolor[rgb]{1.00,0.00,0.50}{2}. The Hamiltonian in Eq.(\ref{KM}) then
becomes%
\begin{equation}
\hat{H}_{\mathrm{KM}}=-\mathit{T}\sum_{\left \langle i,j\right \rangle }\hat
{c}_{i}^{\dagger}\epsilon_{ij}\hat{c}_{j}-\mathit{T}^{\prime}\sum
\limits_{\left \langle \left \langle {i,j}\right \rangle \right \rangle }%
e^{i\phi_{ij}}\hat{c}_{i}^{\dagger}\epsilon_{ij}\hat{c}_{j}-\mu \sum_{i}\hat
{c}_{i}^{\dagger}\hat{c}_{i}.
\end{equation}
Here, $\epsilon_{ij}$ is $-1$ where the $\left \langle i,j\right \rangle $ link
and $\left \langle \left \langle {i,j}\right \rangle \right \rangle $ link cross
the red string which connects two fluxes shown in
Fig.\textcolor[rgb]{1.00,0.00,0.50}{2}, and otherwise, $\epsilon_{ij}$ is $1$.\

\begin{figure}[h]
\centering \scalebox{0.38
}{\includegraphics* [0.8in,0.0in][9.1in,3.8in]{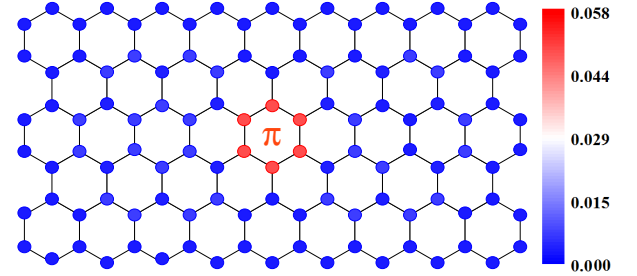}}\caption{The
illustration of the particle density distribution of fermionic zero modes when
two fluxes are well separated. The color of dots denotes the probability of
the particle on the lattice sites. The related parameters in Eq.(1) are chosen
as $\mathit{T}=1.0$, $\mathit{T}^{\prime}=0.1$.}%
\end{figure}By using the exact diagonalization numerical approach on a
$72\times72$ lattice, we find that there exists fermionic zero mode around
each $\pi$-flux. See the particle density around a $\pi$-flux in
Fig.\textcolor[rgb]{1.00,0.00,0.50}{3}. One can see that the particle density
is mainly localized around the $\pi$-flux. The length-scale of the
wave-function of the zero mode is $\xi \sim v_{F}/\Delta_{f}$ where
$v_{F}=3a\mathit{T}/2$ is the Fermi velocity. When there are two fluxes
nearby, the inter-flux quantum tunneling effect occurs and the two zero modes
couple. \begin{figure}[h]
\scalebox{0.52}{\includegraphics* [0.12in,0.1in][10.5in,5.2in]{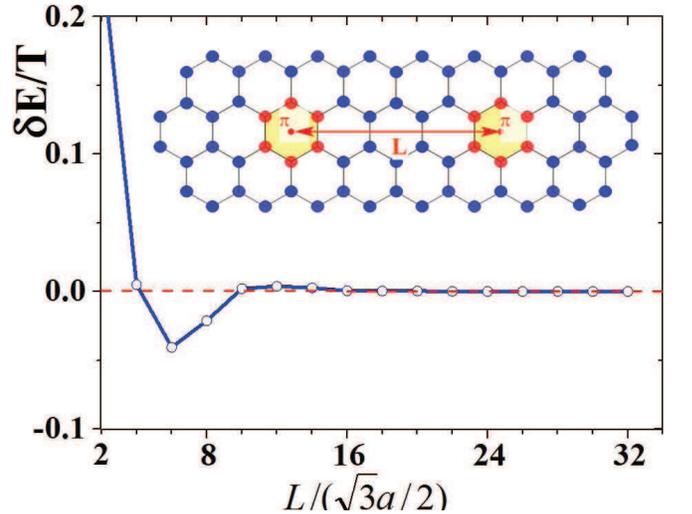}}\caption{The
energy splitting $\delta E$ versus the flux-distance $L$ for the case of
$\mathit{T}=1.0$, $\mathit{T}^{\prime}=0.1$. $a$ is honeycomb lattice
constant. The inset illustrates two $\pi$-flux with the flux-distance $L$.}%
\end{figure}As shown in Fig.\textcolor[rgb]{1.00,0.00,0.50}{4}, the energy
splitting $\delta E$ between two energy levels versus the flux-distance $L$,
oscillates and decreases exponentially. When two fluxes are well separated,
the quantum tunneling effect can be ignored and we have two quantum states
with zero energy. On the other hand, for the small $L$, the coupling between
zero modes becomes stronger and the energy splitting can not be neglected.

\subsection{Mid-gap states of the Haldane model with triangular
flux-superlattice}

We next study the TI with a triangular flux-superlattice.

At the first step, by using the numerical calculations, we obtain the DOS for
the Haldane model with a triangular flux-superlattice. See numerical results
in Fig.\textcolor[rgb]{1.00,0.00,0.50}{5}(b) with flux-superlattice constant
$L=3\sqrt{3}a$ for the case of $\mathit{T}=1.0$, $\mathit{T}^{\prime}=0.1$.
From Fig.\textcolor[rgb]{1.00,0.00,0.50}{5}(b), one may see that the mid-gap
bands appear. In Fig.\textcolor[rgb]{1.00,0.00,0.50}{5}(c), the mid-gap bands
in Fig.\textcolor[rgb]{1.00,0.00,0.50}{5}(b) are zoomed in. Now, we find that
there exist four points with van Hove singularity, and the mid-gap bands also
have an energy gap. \begin{figure}[h]
\includegraphics[width=0.52\textwidth]{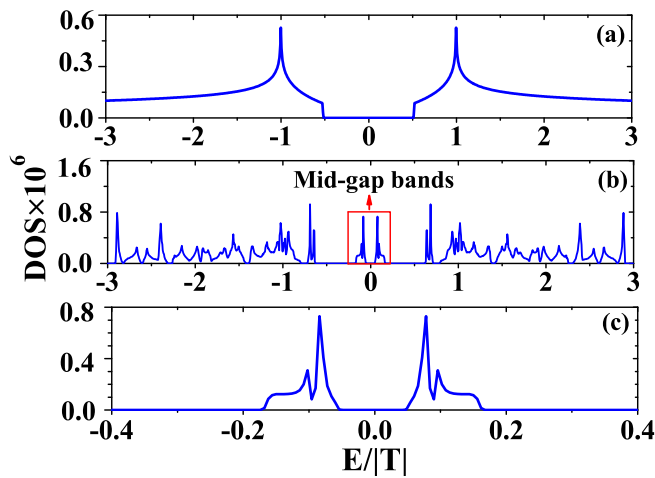}\caption{(a) The DOS of
Haldane model without flux-superlattice for the case of $\mathit{T}=1.0$,
$\mathit{T}^{\prime}=0.1$. (b) The DOS of TI with triangular
flux-superlattice: the mid-gapped states are induced by the flux-superlattice
with flux-superlattice constant $L=3\sqrt{3}a$ for the case of $\mathit{T}%
=1.0$, $\mathit{T}^{\prime}=0.1$. (c) The illustration of details of the
mid-gap band in\ (b).}%
\end{figure}

At second step, to illustrate the topological properties of the mid-gap states
induced by the flux-superlattice of the parent TI, we study the edge states of
the TI with flux-superlattice for the case of $\mathit{T}=1.0$, $\mathit{T}%
^{\prime}=0.1.$ We put the Haldane model with a finite flux-superlattice along
x-direction on a torus. See the illustration in
Fig.\textcolor[rgb]{1.00,0.00,0.50}{6}(a). The parent topological insulator
has the periodic boundary condition along both x-direction and y-direction.
While the flux-superlattice has periodic boundary condition along y-direction
and open boundary condition along x-direction. See the numerical results in
Fig.\textcolor[rgb]{1.00,0.00,0.50}{6}(b). There exist the edge states along
the boundaries of the flux-superlattice. In addition, we plot the particle
density distribution of the edge states in
Fig.\textcolor[rgb]{1.00,0.00,0.50}{6}(a). The existence of the gapless zero
modes on the boundary of the flux-superlattice indicate the mid-gap system is
indeed an induced "topological insulator" on the parent TI.

\begin{figure}[ptb]
\scalebox{0.50}{\includegraphics* [0.0in,0.0in][9.0in,3.9in]{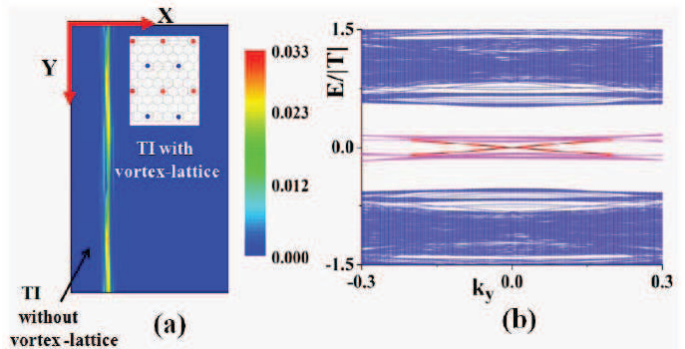}}\caption{(a)
The illustration of particle density of edge states along the boundaries of
flux-superlattice. The flux-superlattice constant is $3\sqrt{3}a$. Related
parameter is chosen as $\mathit{T}^{\prime}/\mathit{T}=0.1$; (b) The edge
states of the composite system, of which the parent Haldane model has periodic
boundary along both x and y directions, and while the flux-superlattice has
periodic boundary along y direction and open boundary condition along x
direction. Parameters (flux-superlattice constant, $\mathit{T}^{\prime}$, and
$\mathit{T}$) are the same as given in (a).}%
\label{Fig5}%
\end{figure}

\section{Effective flux-superlattice model for the mid-gap states}

In this section, we will write down an effective tight-binding model to
describe the mid-gap states.

\subsection{Effective flux-superlattice model}

The quantum states of the fermionic zero mode around a $\pi$-flux can be
formally described in terms of the fermion Fock states $\left \{  \left \vert
0\right \rangle ,\left \vert 1\right \rangle \right \}  $. Here, $\left \vert
0\right \rangle ,\left \vert 1\right \rangle $ denote the un-occupied state and
the occupied state, respectively. These quantum states are localized around
the flux within a length-scale $\xi \sim v_{F}/\Delta_{f}$. For the case of
$\xi<L$, we can consider each flux as an isolated "atom" with localized
electronic states and use the effective tight-binding model to describe these
quantum states on the fluxes.

Now, we superpose the localized states to obtain the sets of Wannier wave
functions $w(R)$ with $R$ denoting the position of the flux. Due to the
formation of the triangular flux-superlattice, both the
quantum-tunneling-strength $\delta E$ between NN fluxes and the
quantum-tunneling-strength $\delta E^{\prime}$ between NNN fluxes exist. The
ratio $\delta E^{\prime}/\delta E$ versus NN flux-superlattice constant $L$ is
shown in Fig.\textcolor[rgb]{1.00,0.00,0.50}{7}. Owing to the inter-flux
quantum tunneling effect, we get energy splitting $\left \vert \delta
E_{RR^{\prime}}\right \vert $ which is just the particle's hopping amplitude
$\left \vert t_{RR^{\prime}}\right \vert $ between two localized states on two
$\pi$-fluxes at $R$ and $R^{\prime}$. Then the effective tight-binding model
of the two fermionic zero modes takes the form of $t_{RR^{\prime}}\left(
\hat{\alpha}_{R}^{\dagger}\hat{\alpha}_{R^{\prime}}+h.c.\right)  $ with
$\left \vert t_{RR^{\prime}}\right \vert =\left \vert \delta E_{RR^{\prime}%
}\right \vert $. With considering the NN and NNN hoppings, the effective
tight-binding model of the localized states around the triangular
flux-superlattice is given by
\begin{equation}
\hat{H}_{\mathrm{VL}}=-\sum_{\left \langle R,R^{\prime}\right \rangle
}t_{RR^{\prime}}\hat{\alpha}_{R}^{\dagger}\hat{\alpha}_{R^{\prime}}%
-\sum_{\left \langle \left \langle R,R^{\prime}\right \rangle \right \rangle
}t_{RR^{\prime}}^{\prime}\hat{\alpha}_{R}^{\dagger}\hat{\alpha}_{R^{\prime}%
},\label{ve}%
\end{equation}
where $\hat{\alpha}_{R}$ is the fermionic annihilation operator of a localized
state on a flux $R$, and $t_{RR^{\prime}}$ ($t_{RR^{\prime}}^{\prime}$) is the
hopping parameter between NN (NNN) sites $R$ and $R^{\prime}$. The NN (NNN)
hopping term is denoted as $\left \vert t_{RR^{\prime}}\right \vert
=t=\left \vert \delta E^{\prime}\right \vert $ ($\left \vert t_{RR^{\prime}%
}^{\prime}\right \vert =\left \vert \delta E^{\prime}\right \vert $). See the
illustration in the inset of Fig.\textcolor[rgb]{1.00,0.00,0.50}{7}. In
particular, owing to the polygon rule (Each fermion gains an accumulated phase
shift $\left \vert \phi \right \vert =(n-2)\pi/2$ encircling around a smallest
\textrm{n}-polygon\cite{rule}), the total phase around each plaquette of the
triangular flux-lattice is $\pm \frac{\pi}{2}$ for fermions. For example, we
may choose the gauge as shown in
Fig.\textcolor[rgb]{1.00,0.00,0.50}{8}.\begin{figure}[ptb]
\includegraphics[width=0.48\textwidth]{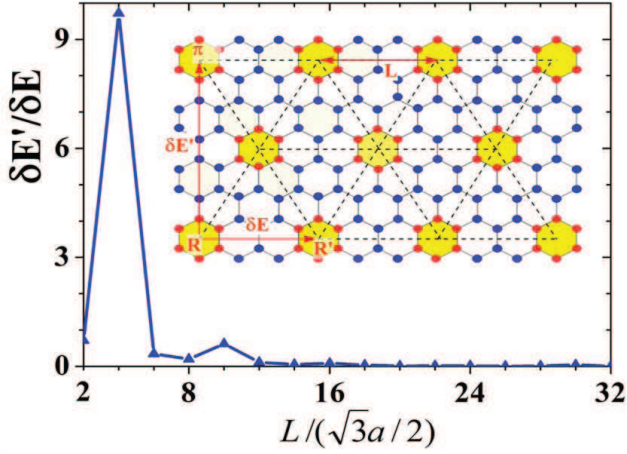}\caption{The ratio between the
NNN quantum tunneling and the NN quantum tunneling ($\delta E/\delta
E^{\prime}$) via flux-superlattice constant $L$, where $\delta E$ versus $L$
is shown in Fig.\textcolor[rgb]{1.00,0.00,0.50}{4}. The inset shows triangular
flux-superlattice. The tunnelings $\delta E$, $\delta E^{\prime}$ and the
flux-superlattice constant $L$ are all indicated in the inset. }%
\end{figure}

\begin{figure}[ptb]
\centering \scalebox{0.50
}{\includegraphics* [0.1in,0.1in][9.6in,4.0in]{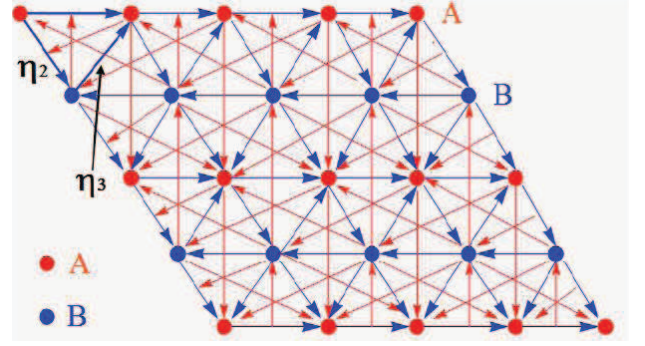}}\caption{The
illustration of the gauge choice. Moving along the arrow (both blue arrow and
red arrow), the fermion acquires a phase shift $\pi/2.$ The accumulated phase
encircling each triangular lattice plaquette anti-clockwise is $\pi/2$.}%
\end{figure}

The ratio between the NN hopping strength $t$ and the NNN hopping strength
$t^{\prime}$ is indeed the same as the ratio between $\left \vert \delta
E\right \vert $ and $\left \vert \delta E^{\prime}\right \vert $. Because people
can tune $\left \vert \delta E^{\prime}\right \vert /\left \vert \delta
E\right \vert $ ($t^{\prime}/t$) by changing the flux-superlattice constant $L$
(see the results in Fig.\textcolor[rgb]{1.00,0.00,0.50}{7}), we may regard the
flux-superlattice model of the mid-gap states as an emergent controllable
topological system.

\subsection{Topological properties}

According to the gauge shown in Fig.\textcolor[rgb]{1.00,0.00,0.50}{8}, we
choose $t_{RR^{\prime}}=\mathrm{i}t$, $t_{RR^{\prime}}^{\prime}=\mathrm{i}%
t^{\prime}$ if fermions hop along directions of arrows. Now we label the
fermionic annihilation operators of the localized states on the two
sub-flux-superlattices by $\hat{A}_{R},$ $\hat{B}_{R}$. See Appendix A for the
detailed formula of Eq.(\ref{ve}) in terms of operators $\hat{A}_{R}$,
$\hat{B}_{R}$. By the Fourier transformation, we obtain the effective
Hamiltonian of the tight-binding model of the TI with triangular
flux-superlattice in momentum space as
\begin{equation}
\hat{H}_{\mathrm{VL}}=\sum_{k}\hat{\Phi}_{k}^{\dagger}\mathcal{H}_{k}\hat
{\Phi}_{k},
\end{equation}
where $\hat{\Phi}_{k}=\left(
\begin{array}
[c]{cc}%
\hat{A}_{k} & \hat{B}_{k}%
\end{array}
\right)  ^{T}$ and $\mathcal{H}_{k}=\left(  \mathbf{h}\cdot \mathbf{\tau
}\right)  $ with $\mathbf{h}=\left(  h_{x},\text{ }h_{y},\text{ }h_{z}\right)
$, and $\mathbf{\tau}$ the Pauli matrix. The three components of $\mathbf{h}$
are
\begin{align}
h^{x} &  =2t\sin \left(  \mathbf{k}\cdot \mathbf{\eta}_{2}\right)  -2t^{\prime
}\sin \left[  \mathbf{k}\cdot \left(  \mathbf{\eta}_{1}+\mathbf{\eta}%
_{3}\right)  \right]  ,\nonumber \\
h^{y} &  =-2t\cos \left(  \mathbf{k}\cdot \mathbf{\eta}_{3}\right)  -2t^{\prime
}\cos \left[  \mathbf{k}\cdot \left(  \mathbf{\eta}_{1}+\mathbf{\eta}%
_{2}\right)  \right]  ,\nonumber \\
h^{z} &  =2t\sin \left(  \mathbf{k}\cdot \mathbf{\eta}_{1}\right)  -2t^{\prime
}\sin \left[  \mathbf{k}\cdot \left(  \mathbf{\eta}_{2}-\mathbf{\eta}%
_{3}\right)  \right]  ,
\end{align}
where $\mathbf{\eta}_{1}=L\left(  1,\text{ }0\right)  $, $\mathbf{\eta}%
_{2}=L(\frac{1}{2},$ $-\frac{\sqrt{3}}{2}),$ $\mathbf{\eta}_{3}=L(\frac{1}%
{2},$ $\frac{\sqrt{3}}{2}).$ In the following, for simplicity, we set the
flux-superlattice constant to be $L\equiv1$. Then energy spectrums of the
effective tight-binding model of the TI with triangular flux-superlattice are
obtained as
\begin{equation}
E_{\mathrm{VL}}(\mathbf{k})=\pm \left \vert \mathbf{h}\right \vert =\pm
\sqrt{(h^{x})^{2}+(h^{y})^{2}+(h^{z})^{2}}.
\end{equation}

The topological quantum phase transition occurs when the band gap closes
$E_{\mathrm{VL}}(\mathbf{k})=0$. See
Fig.\textcolor[rgb]{1.00,0.00,0.50}{9}(a). We find that there exists a quantum
critical point at $t^{\prime}/t=1$ that separates two quantum phases,
$0<t^{\prime}/t<1,$ $t^{\prime}/t>1$. To characterize the two quantum phases,
we introduce the Chern number\cite{dj}
\begin{equation}
C=\frac{1}{4\pi}\int \text{\textrm{tr}}(\mathbf{n\cdot}\frac{\partial
\mathbf{n}}{\partial k_{x}}\times \frac{\partial \mathbf{n}}{\partial k_{y}%
})d^{2}\mathbf{k},
\end{equation}
with $\mathbf{n}=\mathbf{h/}\left \vert \mathbf{h}\right \vert $. According to
the Chern number, we find that in the region of $0<t^{\prime}/t<1,$ the Chern
number is $1$, and\textrm{ }in the region of $t^{\prime}/t>1,$ the Chern
number is $-3$. The tight-binding model of the TI with triangular
flux-superlattice always has nontrivial topological properties. As a result,
we call it \emph{topological} flux-superlattice model that can describe the
mid-gap states of the Haldane model with triangular
flux-superlattice.\begin{figure}[ptb]
\scalebox{0.47}{\includegraphics* [0.00in,0.0in][10.5in,5.5in]{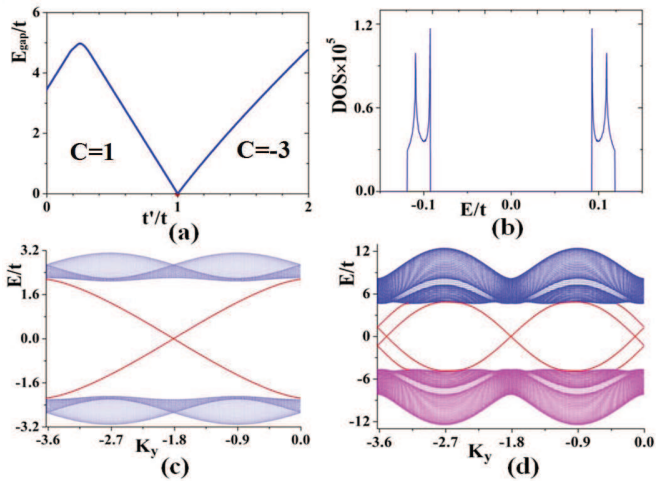}}\caption{(a)
The topological phases versus $t^{\prime}/t$; (b) The density of state of the
free flux-lattice model. The parameters are chosen as $t/\mathit{T}=0.04,$
$t^{\prime}/\mathit{T}=0.014$; (c) The edge states of the tight-binding model
of flux-superlattice for the case of $t^{\prime}=0.1t,$ in which the
Chern-number is $C=1$;\ (d) The edge states of the tight-binding model of
flux-superlattice for the case of $t^{\prime}=3.1t,$ in which the Chern-number
is $C=-3$.}%
\end{figure}

Next, we study the topological properties in different phases. In
Fig.\textcolor[rgb]{1.00,0.00,0.50}{9}(b) we show the DOS of the
flux-superlattice model for the case of $t/\mathit{T}=0.04,$ $t^{\prime
}/\mathit{T}=0.014$. From the DOS, we can see that the effective
flux-superlattice model has a finite energy gap and there exist four points
with van Hove singularity. Fig.\textcolor[rgb]{1.00,0.00,0.50}{9}(c) and
Fig.\textcolor[rgb]{1.00,0.00,0.50}{9}(d) show the edge states for the case of
$t^{\prime}=0.1t$ (the Chern-number is $C=1$) and $t^{\prime}=3.1t$ (the
Chern-number is $C=-3$), respectively.

\subsection{Comparison between the effective flux-superlattice model and
mid-gap system of the parent TI}

The DOS of the Haldane model with triangular flux-superlattice is shown in
Fig.\textcolor[rgb]{1.00,0.00,0.50}{5}(b). Besides the gapped states of the
parent Haldane model, there exist mid-gap states in the energy band gap. It is
obvious that the mid-gap states are induced by the flux-superlattice. The
mid-gap states have an energy gap and there also exist four points with van
Hove singularity. In addition, the gapless edge states of the mid-gap states
are shown in Fig.\textcolor[rgb]{1.00,0.00,0.50}{6}(b). Correspondingly, using
parameters $t$, $t^{\prime}$ derived from the flux-superlattice for the case
of $L=3\sqrt{3}a$, we also get the gapless edge states of the effective
flux-superlattice model, which are similar to those given in Fig.\textcolor[rgb]{1.00,0.00,0.50}{6}(b).

Hence, the low energy physics of the Haldane model with flux-superlattice can
be described by an effective flux-superlattice model. The effective
flux-superlattice model shows nontrivial topological properties, including
nontrivial topological invariant, gapless edge states. In this sense, the
effective flux-superlattice model is really an emergent "topological
insulator" on the parent TIs. The situation is much different from the TI with
a flux-line discussed in Ref.\cite{Assaad}, of which the mid-gap states of a
TI with a flux-line has no nontrivial topological properties.

\section{Conclusions and discussions}

In this paper we mainly studied the Haldane model with triangular
flux-superlattice. We found that there exist the mid-gap states with
nontrivial topological properties, including the nonzero Chern number and the
gapless edge states. We wrote down an effective tight-binding model (the
effective flux-superlattice model) to describe the mid-gap states. Using
similar approach, we studied the Haldane model with other types of
flux-superlattices such as the square flux-superlattice and honeycomb
flux-superlattice, and find similar topological properties. In this sense, the
topological mid-gap states always exist in a TI with flux-superlattices.

In addition, we need to point out that the Kane-Mele model\cite{kane2} or the
spinful Haldane model\cite{he} with flux-superlattice exhibit similar
topological features. Due to the spin degree of freedom, a $\pi$-flux on the
these models traps two zero models. The overlap of zero modes also gives rise
to a topological mid-gap system inside the band gap of parent TIs. When one
considers the interaction between the two-component fermions, the topological
mid-gap system may lead to quite different physics consequences. These issues
are beyond the discussion in this paper and will be studied elsewhere.

\begin{acknowledgments}
This work is supported by NSFC Grant No. 11174035, National Basic Research
Program of China (973 Program) under the grant No. 2011CB921803, 2012CB921704.
\end{acknowledgments}

\appendix

\section{Effective flux-superlattice model of the TI with triangular
flux-superlattice}

The Hamiltonian for the effective flux-superlattice model of the TI with
triangular flux-superlattice is given by%
\begin{equation}
\hat{H}=\hat{H}_{\mathrm{NN}}+\hat{H}_{\mathrm{NNN}},
\end{equation}
where $\hat{H}_{\mathrm{NN}}$ ($\hat{H}_{\mathrm{NNN}}$) denotes the NN (NNN)
hopping term, respectively. The gauge is chosen as shown
Fig.\textcolor[rgb]{1.00,0.00,0.50}{8}. The detailed formulas of $\hat
{H}_{\mathrm{NN}}$ and $\hat{H}_{\mathrm{NNN}}$ are explicitly written as
\begin{align}
\hat{H}_{\mathrm{NN}} &  =\hat{H}_{\mathrm{NN}}^{\mathrm{A}}+\hat
{H}_{\mathrm{NN}}^{\mathrm{B}},\\
\hat{H}_{\mathrm{NNN}} &  =\hat{H}_{\mathrm{NNN}}^{\mathrm{A}}+\hat
{H}_{\mathrm{NNN}}^{\mathrm{B}},
\end{align}
with \begin{widetext}%
\begin{equation}
\hat{H}_{\mathrm{NN}}^{\mathrm{A}}=it\sum_{R}\left(  -A_{R}^{\dagger
}A_{R+\mathbf{\eta}_{1}}+A_{R}^{\dagger}B_{R+\mathbf{\eta}_{3}}+A_{R}%
^{\dagger}B_{R-\mathbf{\eta}_{2}}+A_{R}^{\dagger}A_{R-\mathbf{\eta}_{1}}%
+A_{R}^{\dagger}B_{R-\mathbf{\eta}_{3}}-A_{R}^{\dagger}B_{R+\mathbf{\eta}_{2}%
}\right)
\end{equation}%
\begin{equation}
\hat{H}_{\mathrm{NN}}^{\mathrm{B}}=it\sum_{R}\left(  B_{R}^{\dagger
}B_{R+\mathbf{\eta}_{1}}-B_{R}^{\dagger}A_{R+\mathbf{\eta}_{3}}+B_{R}%
^{\dagger}A_{R-\mathbf{\eta}_{2}}-B_{R}^{\dagger}B_{R-\mathbf{\eta}_{1}}%
-B_{R}^{\dagger}A_{R-\mathbf{\eta}_{3}}-B_{R}^{\dagger}A_{R+\mathbf{\eta}_{2}%
}\right)
\end{equation}%
\begin{equation}
\hat{H}_{\mathrm{NNN}}^{\mathrm{A}}=it^{\prime}\sum_{R}\left(  A_{R}^{\dagger
}B_{R+\mathbf{\eta}_{1}+\mathbf{\eta}_{3}}-A_{R}^{\dagger}A_{R-\mathbf{\eta
}_{2}+\mathbf{\eta}_{3}}+A_{R}^{\dagger}B_{R-\mathbf{\eta}_{1}-\mathbf{\eta
}_{2}}-A_{R}^{\dagger}B_{R-\mathbf{\eta}_{1}-\mathbf{\eta}_{3}}+A_{R}%
^{\dagger}A_{R+\mathbf{\eta}_{2}-\mathbf{\eta}_{3}}+A_{R}^{\dagger
}B_{R+\mathbf{\eta}_{1}+\mathbf{\eta}_{2}}\right)
\end{equation}%
\begin{equation}
\hat{H}_{\mathrm{NNN}}^{\mathrm{B}}=it^{\prime}\sum_{R}\left(  B_{R}^{\dagger
}A_{R+\mathbf{\eta}_{1}+\mathbf{\eta}_{3}}+B_{R}^{\dagger}B_{R-\mathbf{\eta
}_{2}+\mathbf{\eta}_{3}}-B_{R}^{\dagger}A_{R-\mathbf{\eta}_{1}-\mathbf{\eta
}_{2}}-B_{R}^{\dagger}A_{R-\mathbf{\eta}_{1}-\mathbf{\eta}_{3}}-B_{R}%
^{\dagger}B_{R+\mathbf{\eta}_{2}-\mathbf{\eta}_{3}}-B_{R}^{\dagger
}A_{R+\mathbf{\eta}_{1}+\mathbf{\eta}_{2}}\right)
\end{equation}
\end{widetext}

\end{document}